\documentclass{article}

\usepackage[english]{babel}

\usepackage[letterpaper,top=2cm,bottom=2cm,left=3cm,right=3cm,marginparwidth=1.75cm]{geometry}

\usepackage{amsmath}
\usepackage{graphicx}
\usepackage[colorlinks=true, allcolors=blue]{hyperref}
\usepackage{xcolor}
\usepackage{float}
\usepackage[round,authoryear]{natbib}

\title{Beyond Prediction: Longitudinal Reasoning in EHR-Integrated Clinical AI}
\author{
  Irene Yi$^1$, Grace Brown$^{1,*}$, Sufian Aldogom$^{2,*}$, Nathan Roll$^1$, Eric J. Basile DO$^5$, \\
  Pamela M. Resnikoff MD MPH$^4$$^,$$^5$, Isaac Gutterman$^3$,\\
Oscar Schiff$^3$, Keira Salata$^3$, Benjamin Mujkic$^3$, Ammar Ahmed MD$^5$ \\
  $^1$Stanford University \\
  $^2$Massachusetts Institute of Technology \\
  $^3$Harvard University \\
  $^4$Scripps Mercy Hospital San Diego Pulmonary Medicine \\
  $^5$Aurevia MD \\
  \small{$^*$Primary authors}
}

\begin{document}
\maketitle

\begin{abstract}
We present a structured analysis of how contemporary clinical AI systems integrate electronic health record (EHR) data and the extent to which they support longitudinal clinical reasoning. Drawing on a curated corpus of clinical natural language processing (NLP) and EHR-integrated systems, we develop a coding framework that captures both technical integration strategies and reasoning-relevant representational features, such as trajectory modeling, cross-encounter synthesis, longitudinal analysis, and absence reasoning. We also elicited the experiences of three physicians in their EHR use, including what strengths and weaknesses they found with their institution’s current EHR system(s). Our analysis shows that while many systems incorporate EHR data, they predominantly operate on encounter-level or aggregated representations, with limited support for explicit temporal reasoning across patient histories. Reasoning-relevant structures are inconsistently represented, and evaluation paradigms remain largely focused on predictive performance instead of longitudinal interpretability. We argue that current approaches treat EHR data as a static input rather than a substrate for ongoing clinical reasoning, and we outline a framework for understanding how future systems might more effectively align with the temporal and interpretive structure of clinical practice.
\end{abstract}

\section{Introduction}

Electronic health records (EHRs) have become the central infrastructure of contemporary clinical practice, organizing patient information across temporal and institutional contexts. As a result, they also serve as the primary substrate for the development of clinical artificial intelligence systems. Over the past decade, advances in machine learning and natural language processing have enabled increasingly sophisticated models that operate over EHR data, supporting tasks like diagnosis prediction and clinical documentation. These developments have positioned EHR-integrated AI as a key site of innovation in healthcare. More broadly, the growing integration of AI into clinical care reflects a shift toward augmenting physician expertise, with intelligent systems increasingly viewed as tools that can enhance clinical decision-making and reduce administrative burden when appropriately integrated into existing systems \citep{Topol2019, Friedman1997}.

The centrality of EHR as a substrate for AI is the product of federal policy. The 2009 Health Information Technology for Economic and Clinical Health (HITECH) Act, enacted under the American Recovery and Reinvestment Act, tied billions of dollars in Medicare and Medicaid incentives to the ``meaningful use'' of certified EHRs \citep{HITECH2009, Blumenthal2010}. The resulting repository, however, was shaped primarily by administrative, billing, and regulatory requirements rather than by the cumulative, interpretive nature of clinical reasoning. EHRs, therefore, constitute a vast but \emph{forced} substrate for clinical AI, whose data structures often reflect compliance needs and subsequently create a structural mismatch between the organization of the record and the reasoning it is meant to support.

At the same time, clinical reasoning remains fundamentally a temporal and cumulative process. Physicians do not reason within isolated encounters; instead, they construct and revise interpretations of patient state over time, integrating information across visits, tracking trajectories of change, and reinterpreting prior findings in light of new evidence. This process involves the accumulation of information and the selective organization of that information into meaningful structures, such as patterns or absences. While EHR systems provide access to the data underlying this reasoning, they do not necessarily structure that data in ways that align with how clinicians interpret and use it.

This raises a broader question about the relationship between EHR-integrated AI systems and the structure of clinical reasoning. Although many systems incorporate longitudinal data in some form, it remains unclear to what extent they support reasoning processes that depend on temporal continuity, cross-encounter integration, and interpretive flexibility. In particular, there is a need to move beyond evaluating systems solely in terms of predictive performance and toward understanding how they represent and make available the elements of reasoning that clinicians rely on in practice.

In this paper, we take a representational perspective on EHR-integrated AI systems, focusing on how they encode and operate over patient data across time. We argue that the key issue is not simply whether systems use longitudinal data, but how that data is structured in relation to clinical reasoning processes. To address this, we develop a coding framework that captures both technical design features and reasoning-relevant affordances, and apply it to a curated corpus of systems spanning clinical NLP, decision support, and EHR-based machine learning. By systematically characterizing the current design space, we aim to clarify how existing approaches align with, approximate, or diverge from the temporal and interpretive structure of clinical reasoning.

\section{Background}

AI-integrated EHR systems are built for clinical decision support and longitudinal health data processing. A common theme in this area is the use of EHR data as a rich source of information for modeling and prediction. The increasing availability of large-scale electronic health records has fundamentally transformed biomedical research and clinical informatics, enabling population-scale analyses while creating new opportunities and challenges for computational representations of patient state \citep{Jensen2012}. Early work in clinical decision support systems focused on rule-based approaches that leveraged structured data to generate alerts and recommendations at the point of care \citep{Bates2003, Bright2012, Kawamoto2005}. While these systems demonstrated the potential to improve clinical outcomes, doctors also faced issues related to usability and workflow integration, underscoring the difficulty of embedding computational support within complex clinical environments. Subsequent reviews have similarly emphasized that successful clinical decision support depends on careful integration into clinical workflows and user-centered design \citep{Sutton2020}.

More recent work has shifted toward data-driven approaches, particularly machine learning models trained on large-scale EHR datasets. These models have achieved strong performance on a range of predictive tasks, including mortality prediction, hospital readmission, and disease onset \citep{Rajkomar2018, Choi2016}. Many of these approaches explicitly incorporate temporal information, using sequence models such as recurrent neural networks or transformer-based architectures to capture patterns over time. This line of work reflects an increasing recognition that patient data is inherently longitudinal and that temporal dynamics are important for accurate prediction. Advances in deep learning have expanded the range of clinical tasks that can be approached computationally, including diagnosis, prognosis, imaging, multimodal reasoning, and more, further establishing AI as a foundational component of modern healthcare research \citep{Esteva2019}.

However, the incorporation of temporal data in these models does not necessarily entail support for clinical reasoning as it is practiced. In many cases \citep{Rudin2019, Tonekaboni2019, Shortliffe2018}, longitudinal information is encoded in ways that are optimized for predictive performance rather than interpretability or alignment with clinician workflows. For example, sequence models may learn representations of temporal patterns without exposing interpretable trajectories or supporting explicit reasoning across encounters. Similarly, predictive models often operate at fixed prediction points, producing outputs that summarize risk without preserving the underlying temporal structure of patient histories. 

In parallel, advances in clinical natural language processing have focused on extracting and generating information from unstructured clinical notes. Recent work using large language models has demonstrated strong performance on tasks such as summarization or question answering, as well as automated documentation \citep{Singhal2023}. These systems have been increasingly integrated into clinical workflows, particularly through tools that assist with note generation and summarization. While such approaches improve the accessibility and usability of clinical data, they are typically oriented toward encounter-level representations and may not capture relationships across visits or the evolution of patient state over time.

A related line of work has examined the integration of AI systems into EHR platforms and clinical workflows. Studies of real-world deployments emphasize issues such as usability, trust, and alignment with clinician needs \citep{Shah2019, Chenais2023}. These studies highlight the importance of designing systems that fit within existing practices, but they often focus on adoption and user experience rather than the representational structure of the systems themselves. Across these domains, a key limitation is that EHR data is frequently treated as a static input to modeling pipelines rather than as a dynamic substrate for reasoning. Even when systems incorporate longitudinal data, this information is often aggregated or compressed in ways that obscure temporal relationships. As a result, important aspects of clinical reasoning, such as trajectory tracking, cross-encounter synthesis, and the interpretation of absent findings, are not explicitly represented in most systems. This gap has begun to be recognized in recent discussions of explainability and human-AI collaboration, which emphasize the need for systems that align more closely with clinician mental models \citep{DoshiVelez2017, Rudin2019}. Studies examining clinician perspectives on explainable AI suggest that physicians often prioritize contextualized explanations and transparent reasoning pathways \citep{Tonekaboni2019}. However, these efforts typically focus on making individual predictions more interpretable and do not address how systems represent patient data across time or support reasoning processes that unfold longitudinally. 

In light of these limitations, there is a need for a more systematic account of how EHR-integrated AI systems engage with the structure of clinical reasoning. In this paper, we address this need by developing a coding framework that captures reasoning-relevant representational features and applying it to a corpus of existing systems. This approach allows us to move beyond task-specific evaluation and toward a broader understanding of how current systems align with, approximate, or diverge from the temporal and interpretive structure of clinical reasoning in practice.

\section{Research Questions and Objectives}

Taken together, prior work demonstrates substantial progress in applying machine learning and natural language processing to EHR data, but leaves open the question of how these systems engage with the structure of clinical reasoning over time. In particular, while many approaches incorporate longitudinal data, it remains unclear how that data is represented in relation to reasoning processes such as trajectory tracking, cross-encounter integration, and the interpretation of absent or implicit information. This paper asks: How do current EHR-integrated AI systems represent and operationalize longitudinal patient data, and to what extent do these representations align with the structure of clinical reasoning in practice? To address this overarching question, we further examine: Which aspects of clinical reasoning (such as temporal dynamics, cross-encounter synthesis, and implicit or absence-based reasoning) are supported, approximated, or not represented in existing systems?

\section{Methodology}
This study employs a structured literature review with a systematic coding framework to analyze how contemporary clinical AI systems integrate electronic health record (EHR) data and the extent to which they support longitudinal clinical reasoning. The analysis is designed to characterize how systems represent, access, and operationalize patient data over time, with particular attention to reasoning-relevant structures such as trajectories, cross-encounter synthesis, and the treatment of absent or missing information. To accomplish this, we constructed a corpus of papers spanning clinical natural language processing, clinical decision support, and EHR-integrated machine learning systems. Papers were selected according to three criteria: (1) the system explicitly consumes or operates over EHR data, (2) the system performs a clinically meaningful task such as diagnosis support, summarization, or prediction, and (3) the paper provides sufficient methodological detail to allow assessment of how EHR data is structured and used. The goal of corpus construction was to gather a representative sample of current system design patterns across domains and tasks rather than obtain exhaustive coverage of all existing systems. The coding framework for these papers will be explained in section 4.2 below.

\subsection{Physician Perspectives on EHR Use}
To complement the structured analysis of existing systems, we incorporated qualitative reflections from practicing physicians on their experiences using EHR systems in routine clinical work. Three physicians, representing different clinical specialties and institutional contexts, were invited to provide written accounts of their interaction with EHR platforms at their respective institutions. Participants were selected to provide variation in clinical setting and workflow, rather than to serve as a statistically representative sample. Each physician was asked to describe their typical use of the EHR during patient care, including how they access, interpret, and synthesize patient information across encounters. Prompts also encouraged reflection on aspects of the system that support or hinder clinical reasoning, with particular attention to navigating longitudinal patient histories, identifying relevant information, and reconstructing patient trajectories over time. In addition, physicians were asked to identify specific areas where current systems could be improved to better align with their reasoning processes and workflow needs.

These narratives were analyzed qualitatively and used to contextualize the findings from the literature review. The physician reflections provide grounded insight into how EHR systems are experienced in practice and highlight recurring challenges in usability and information organization. In particular, they offer a clinician-centered perspective on issues such as fragmentation of information across encounters, the difficulty of tracking changes over time, and the effort required to reconstruct patient histories from existing documentation.

The inclusion of these perspectives allows us to connect system-level design patterns identified in the literature with lived clinical experience, strengthening the interpretive link between technical implementation and practical use. While limited in scale, these accounts serve as an illustrative bridge between the representational capacities of current systems and the demands of real-world clinical reasoning.

\subsection{Meta-analysis Coding Framework}
To characterize how artificial intelligence systems engage with electronic health record (EHR) data and clinical reasoning processes, each paper in the corpus was independently coded across a standardized set of technical and representational dimensions. The coding framework was developed iteratively through review of the clinical AI, EHR prediction, clinical natural language processing, and clinical decision support literature. Coding was designed to distinguish model architecture and output type, as well as the extent to which systems represent longitudinal patient histories and temporal dynamics and trajectories. The framework is highlighted in Table 1.

\begin{table*}[t][H]
\centering
\caption{Coding framework used to characterize EHR-based AI systems}
\label{tab:ehr_coding_framework}
\small
\begin{tabular}{p{3.6cm} p{11cm}}
\hline
\textbf{Variable} & \textbf{Operational Definition} \\
\hline

Category &
Primary system function. Coded as Prediction, Summarization, Retrieval, Clinical Decision Support (CDS), or Other. \\

Uses Longitudinal Data &
Whether the system incorporates information from multiple encounters, repeated observations, or temporally ordered patient records. (Y/N/Unclear) \\

Persistent Patient State &
Whether the model maintains an evolving representation of an individual patient across encounters or time. (Y/N/Unclear) \\

Temporal Modeling &
Extent to which temporal information is represented. Coded as None, Sequential, Time-Aware, or Continuous-Time. \\

Absence Reasoning &
Whether the system explicitly represents missingness, negative findings, lack of expected events, or absence of observations as evidence. (Y/N/Unclear) \\

Trajectory Reasoning &
Whether the system models patient progression, deterioration, recovery, evolving diagnoses, or changing risk over time. (Y/N/Unclear) \\

Counterfactual Reasoning &
Whether the system estimates hypothetical outcomes, treatment effects, alternative trajectories, or intervention-based changes. (Y/N/Unclear) \\

Cross-Encounter Reasoning &
Whether information from multiple encounters is integrated into a unified patient representation. (Y/N/Unclear) \\

Trajectory Modeling Type &
How patient evolution is represented. Coded as None, Risk Forecasting, Disease Progression, or Multi-State Trajectory. \\

Output Type &
Primary system output. Coded as Risk, Summary, Alert, Recommendation, or Other. \\

LLM-Based &
Whether the system relies primarily on a large language model or foundation model architecture. (Y/N/Unclear) \\

Retrieval-Based &
Whether retrieval mechanisms or retrieval-augmented generation are core components of the system. (Y/N/Unclear) \\

Rule-Based &
Whether symbolic rules, ontologies, or expert systems play a central role in system behavior. (Y/N/Unclear) \\

Patient Representation Level &
Scope of information maintained by the system. Coded as Encounter-Level, Multi-Encounter, Longitudinal Patient State, or Population-Level. \\

Information Compression Level &
Degree of abstraction applied to patient information. Coded as Raw Retrieval, Feature Aggregation, Latent Embedding, Summary Generation, or Recommendation Generation. \\

Longitudinal Fidelity &
Extent to which the original longitudinal patient trajectory remains represented. Coded as Very High, High, Medium, Low, or Very Low. \\

Benchmark vs Deployed &
Whether the paper evaluates a research/benchmark system or reports a deployed clinical implementation. \\

EHR-Native vs General &
Whether the system is designed around EHR data structures, general clinical NLP, general machine learning, or hybrid approaches. \\

Confidence &
Confidence assigned during coding adjudication based on clarity of methods and reporting. (High/Medium/Low) \\

\hline
\end{tabular}
\end{table*}

Each paper was assigned a unique identifier and coded according to bibliographic metadata (title, authors, year, venue, and system name), system category, reasoning characteristics, implementation characteristics, and deployment context. Categories included prediction systems, summarization systems, retrieval systems, clinical decision support systems, and other AI-enabled healthcare applications.

Longitudinal data usage was coded as present when a system incorporated information from multiple patient encounters, repeated observations, or temporally ordered health records. Persistent patient state was coded when a model maintained an evolving representation of a patient across time, as opposed to treating encounters independently. Temporal modeling was categorized as none, implicit, or explicit depending on whether temporal structure was absent, encoded indirectly through sequence order, or represented through timestamps, temporal attention mechanisms, decay functions, continuous-time modeling, or similar approaches.

Reasoning-related variables were coded to capture higher-level representations of patient evolution. Trajectory reasoning was coded when a system modeled clinical progression, disease evolution, future diagnoses, deterioration, or changing risk over time. Cross-encounter reasoning was coded when information from multiple visits or encounters was integrated into a unified inference process. Absence reasoning was coded only when models explicitly represented missingness, negative evidence, or the absence of observations as part of inference. Counterfactual reasoning was coded only when systems estimated hypothetical outcomes, intervention effects, or alternative future trajectories.

Implementation-oriented variables included whether a system was based on large language models, retrieval mechanisms, or explicit rule-based logic. Output types were classified as risk prediction, summary generation, alert generation, recommendation generation, or other outputs. Each paper was additionally categorized according to whether it represented a benchmark/research system or a deployed clinical system, and whether it was EHR-native, general clinical natural language processing, general machine learning, or a hybrid approach.

To improve reproducibility, operational definitions were established before adjudication. In particular, temporal modeling and trajectory reasoning were treated as distinct constructs. Systems could contain explicit temporal representations without performing trajectory reasoning, while trajectory reasoning required modeling clinically meaningful patient evolution. Review articles, surveys, benchmarks, and governance papers were coded according to the properties of the paper itself rather than inheriting characteristics from the systems they described.

Following initial coding, a multi-stage adjudication process was conducted to identify duplicate papers, reconcile conflicting codes, standardize bibliographic metadata, and ensure consistent application of coding definitions across the corpus. This process produced a final audited dataset suitable for descriptive analysis of longitudinal reasoning, temporal representation, retrieval augmentation, and clinical AI system design.

\section{Results}
\subsection{Meta-Analysis Results}
The final corpus consisted of 85 unique publications spanning clinical decision support, clinical natural language processing, predictive modeling, EHR infrastructure, and foundation-model-based healthcare AI systems. Across the corpus, relatively few systems maintained explicit representations of longitudinal patient state, while the majority operated on aggregated features, encounter-level information, or population-level abstractions.

Analysis of patient representation revealed that most systems were coded as population-level representations (50/85, 58.8\%), indicating that they primarily operated on cohorts, aggregate data structures, governance frameworks, or generalized predictive pipelines. Encounter-level representations accounted for 18 systems (21.2\%), while only 17 systems (20.0\%) were coded as maintaining longitudinal patient-state representations spanning multiple encounters and evolving clinical histories.

\begin{figure}[H]
    \centering
    \includegraphics[width=1\linewidth]{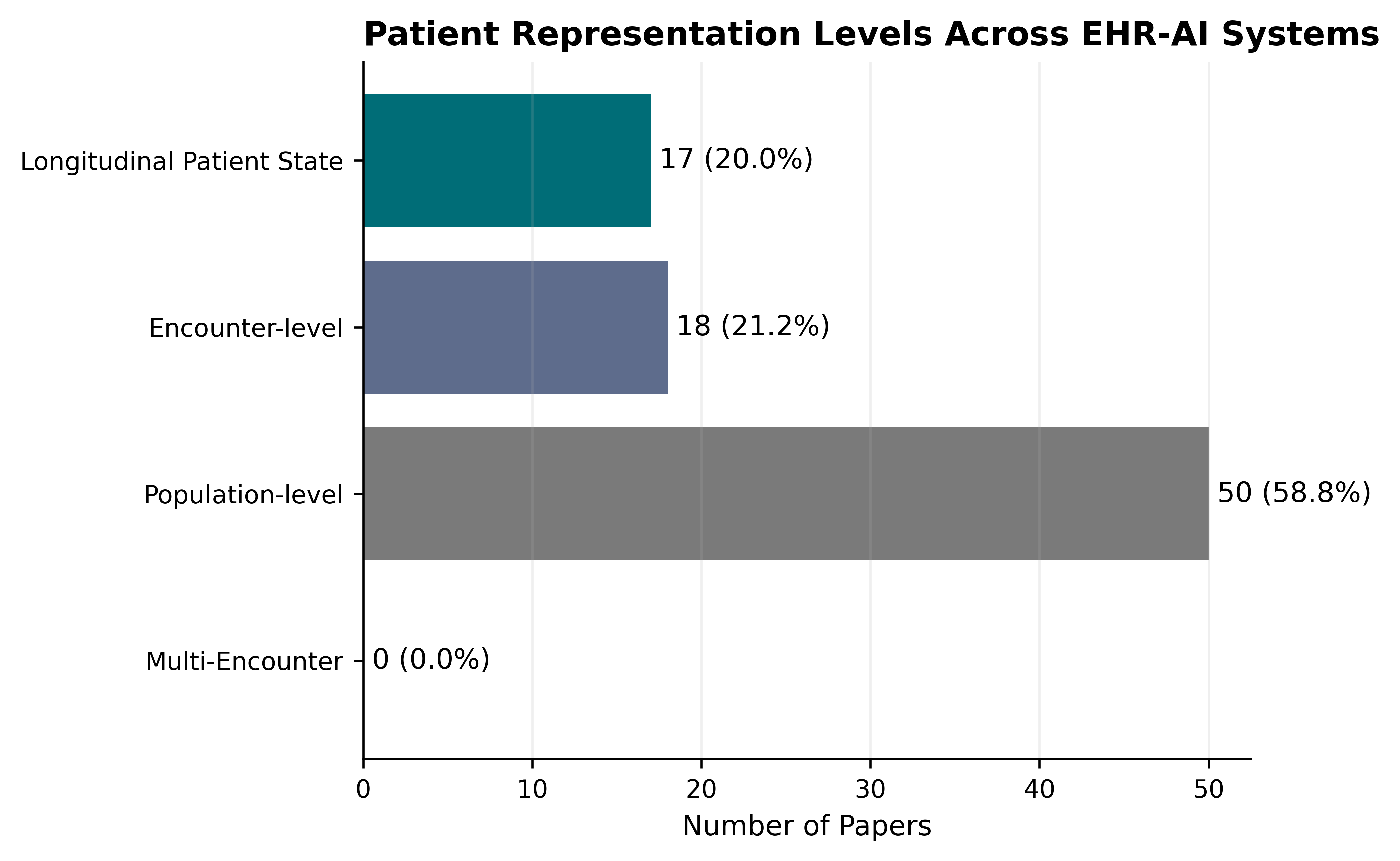}
    \caption{Patient representation levels in EHR-AI systems}
    \label{Figure 2. Patient representation levels in EHR-AI systems}
\end{figure}

This distribution shows that much of the EHR-AI literature is organized around aggregate modeling, system design, benchmarking, governance, or encounter-specific processing rather than sustained patient-level representation. The significance of this finding is that the dominant unit of analysis in the literature is often not the longitudinal patient, even when the underlying data source is the EHR.

Temporal representation was similarly limited. Most systems did not explicitly model temporal structure (68/85, 80.0\%). Only 15 systems (17.6\%) incorporated time-aware representations, while 2 systems (2.4\%) used sequential temporal modeling without explicit time-aware mechanisms. Although many studies utilized longitudinal EHR data, comparatively few maintained temporally explicit representations throughout the modeling process.

Information compression emerged as a dominant characteristic of contemporary EHR-based AI systems. More than half of the corpus (48/85, 56.5\%) relied primarily on feature aggregation approaches, transforming complex patient records into structured variables or extracted concepts. An additional 18 systems (21.2\%) compressed patient histories into latent embeddings. Recommendation-generating systems accounted for 15 papers (17.6\%), while only 2 systems (2.4\%) primarily functioned as raw retrieval infrastructures preserving underlying patient information. Summary-generation systems represented a similarly small proportion of the corpus (2.4\%).

\begin{figure}[H]
    \centering
    \includegraphics[width=1\linewidth]{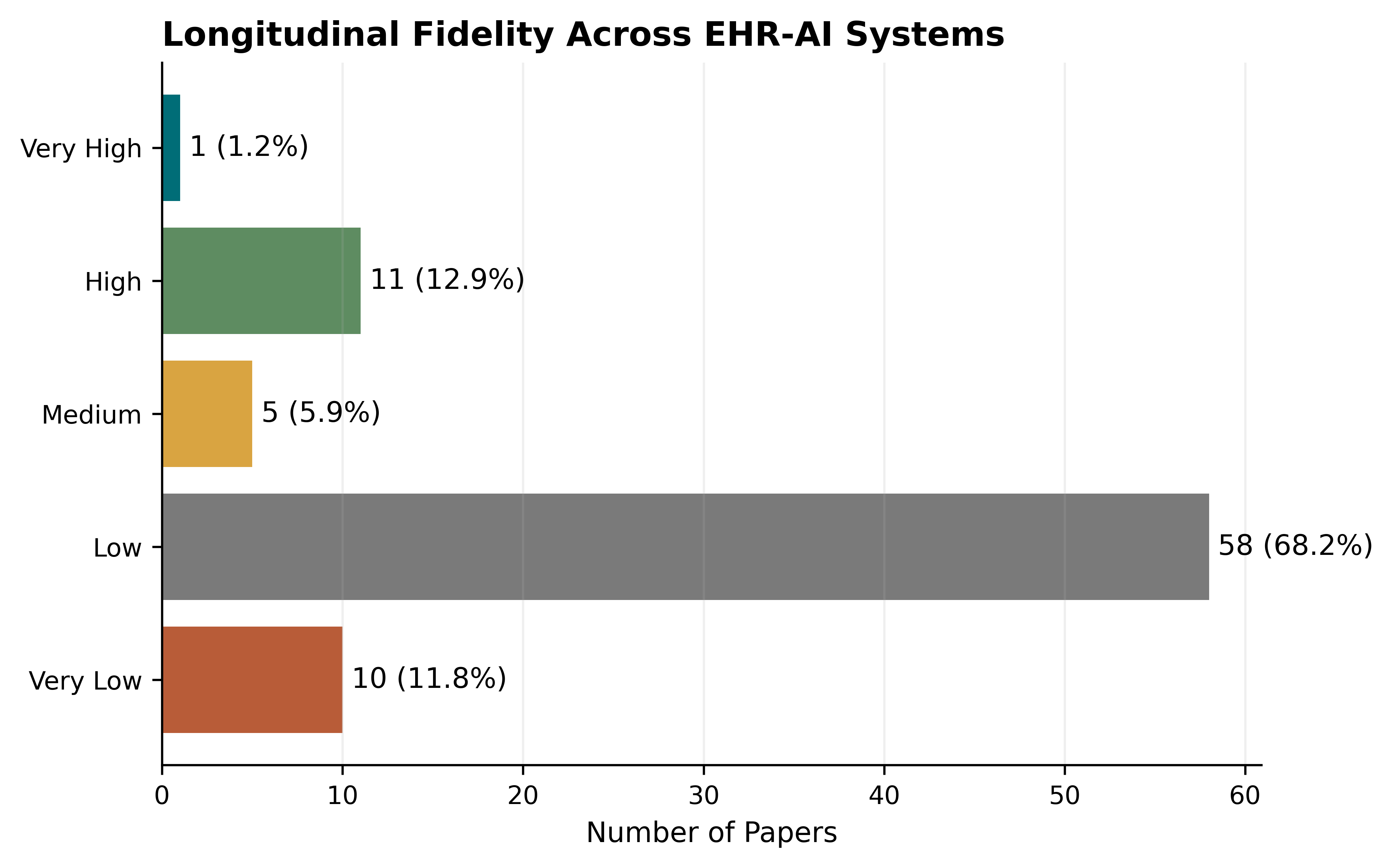}
    \caption{Longitudinal fidelity across EHR-AI systems}
    \label{Figure 1. Longitudinal fidelity across EHR-AI systems}
\end{figure}

The most striking finding concerned longitudinal fidelity. The majority of systems (58/85, 68.2\%) exhibited low longitudinal fidelity, indicating that substantial portions of the original longitudinal patient narrative were compressed or abstracted during processing. Only 11 systems (12.9\%) achieved high longitudinal fidelity, while a single system (1.2\%) was coded as preserving very high longitudinal fidelity. Ten systems (11.8\%) demonstrated very low longitudinal fidelity, typically transforming patient information into recommendations, generated responses, or highly compressed outputs. 

This suggests that although EHR-based AI systems often rely on patient data, relatively few preserve the patient’s longitudinal clinical course as an explicit representational object. The finding supports the paper’s central claim that access to EHR data should not be equated with preservation of longitudinal patient state.

\begin{figure}[H]
    \centering
    \includegraphics[width=1\linewidth]{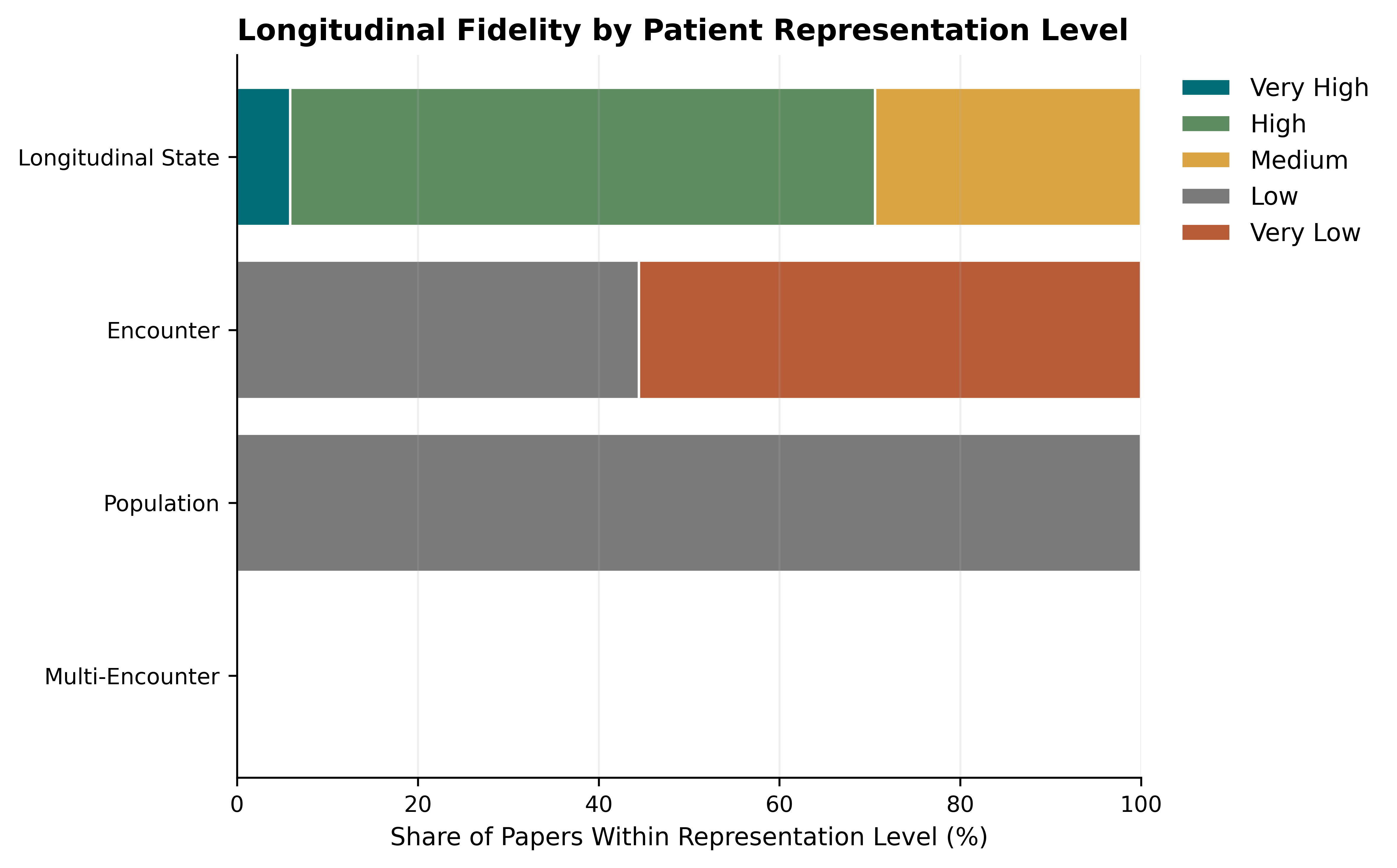}
    \caption{Patient representation level by longitudinal fidelity}
    \label{Figure 3. Patient representation level by longitudinal fidelity}
\end{figure}

The relationship between representation level and longitudinal fidelity was highly structured. Longitudinal patient-state systems accounted for nearly all high-fidelity representations, including 11 high-fidelity papers, 5 medium-fidelity papers, and the single very-high-fidelity paper. By contrast, all encounter-level systems were coded as either low or very low fidelity, and population-level papers were uniformly low fidelity. This pattern indicates that high longitudinal fidelity is not a general property of EHR-based AI, but is concentrated in a relatively small subset of models that explicitly maintain patient-state representations over time. 

Figure 3 also identifies a representational divide within the EHR-AI literature. Systems that maintain explicit longitudinal patient-state representations consistently preserve more of the temporal and contextual structure of clinical histories, whereas encounter-level and population-level systems tend to abstract patient information into increasingly compressed representations. This suggests that longitudinal fidelity is fundamentally linked to representational architecture, highlighting patient-state modeling as a critical but relatively uncommon paradigm within contemporary EHR-based AI systems.

Taken together, these findings suggest that while many contemporary AI systems are built upon EHR data, relatively few preserve longitudinal patient trajectories as first-class representations. Instead, most systems transform patient histories into progressively more compressed representations, including aggregated features, latent embeddings, risk scores, alerts, recommendations, or generated outputs. This pattern suggests a representational shift in which rich longitudinal clinical histories are increasingly abstracted as they move through AI pipelines. Consequently, the ability of AI systems to access longitudinal EHR data should not be conflated with preservation of longitudinal patient state. Across the corpus, explicit maintenance of evolving patient trajectories remained comparatively uncommon relative to systems focused on prediction, aggregation, summarization, or recommendation generation.

\subsection{Physician Reflections}
To complement the literature review, three physicians with experience using multiple EHR systems provided written reflections on how EHR design influences clinical workflow, information retrieval, and patient care. Although the reflections varied in specialty and practice environment, several consistent themes emerged.

\subsubsection{Information Accessibility}
All physicians emphasized that modern EHRs contain large amounts of clinical information but often make it difficult to identify what is most relevant. While systems such as Epic were generally viewed as more usable than alternatives, clinicians consistently described substantial effort devoted to locating and synthesizing information scattered across notes, laboratory results, imaging studies, and prior encounters. One physician noted that the EHR provides access to extensive information but offers little assistance in distinguishing clinically important information from irrelevant details, leaving physicians responsible for performing this synthesis themselves under significant time pressure. Another described the challenge of assembling a coherent longitudinal picture from information distributed across multiple sections of the chart. Together, these observations suggest that current EHRs function primarily as repositories of information rather than systems that actively support longitudinal clinical understanding.

\subsubsection{Longitudinal Context}
A second recurring theme was the difficulty of reconstructing a patient's clinical trajectory across time. Physicians described spending substantial time reviewing prior notes, laboratory trends, imaging results, and specialist evaluations before encounters, particularly for medically complex patients. Despite these efforts, clinicians frequently reported uncertainty regarding whether they had successfully identified all clinically relevant information. One physician noted that understanding complex patients often requires multiple visits, not because information is absent from the record, but because the record does not effectively surface how clinical events, decisions, and interpretations evolved over time. Another emphasized that reviewing trends and prior data becomes increasingly cumbersome when EHR systems lack intuitive mechanisms for displaying longitudinal information. 

\subsubsection{Documentation Priorities}
Physicians repeatedly described tension between documentation requirements and clinical reasoning. One respondent argued that the structure of contemporary EHR documentation reflects billing and regulatory requirements more strongly than the needs of clinical decision-making. Template-driven documentation and copy-forward practices were viewed as contributing to "note bloat," making it increasingly difficult to identify clinically meaningful information within notes. Several physicians reported that even when reviewing their own prior documentation, they were often unable to reconstruct the reasoning that led to a particular clinical assessment or decision. 

\subsubsection{Cognitive Burden}
All respondents described EHR use as a substantial contributor to clinical workload. Documentation, order entry, chart review, and navigation were repeatedly cited as major consumers of physician time. Physicians reported spending considerable portions of patient encounters interacting with the EHR rather than directly engaging with patients. One physician described a persistent tension between maintaining eye contact and documenting information in real time, while another characterized chart review as a major source of cognitive overhead both before and after patient encounters. These reflections suggest that inefficiencies in information organization and retrieval may impose cognitive costs that extend beyond simple time expenditure.

\subsubsection{Information Synthesis}
Despite differences in systems and practice settings, physicians consistently expressed a desire for tools that would help organize and interpret existing information rather than simply provide access to larger quantities of data. Respondents generally viewed features such as search functions, longitudinal trend visualization, integrated communication tools, and AI-assisted summarization positively when they reduced the effort required to locate or organize information. However, physicians also expressed concern that summary-generating technologies could obscure important details, introduce inaccuracies, or further distance clinicians from the underlying evidence. This tension suggests that future EHR-integrated AI systems may be most valuable when they support information synthesis while preserving transparency and traceability to source data.

Across all three narratives, physicians consistently described a gap between the volume of information available within modern EHRs and the degree of clinical understanding that can be extracted from that information under real-world time constraints. Although contemporary EHRs successfully capture and store large amounts of clinical data, clinicians reported persistent difficulties reconstructing patient trajectories, identifying relevant information, and understanding how prior clinical conclusions were reached. These themes closely mirror the findings of the literature review, which demonstrated that many EHR-integrated AI systems consume longitudinal clinical data while providing limited support for preserving or reconstructing longitudinal patient state. Taken together, the physician reflections suggest that challenges of information organization, trajectory reconstruction, and representational fidelity are not merely technical concerns but directly influence clinical workflow and patient care.

\section{Discussion}
This review examined how contemporary AI systems represent and utilize information derived from electronic health records. Across the corpus, a consistent pattern emerged: EHR integration does not necessarily imply support for longitudinal patient representation. Although many systems incorporated EHR data as input, relatively few maintained explicit representations of evolving patient state over time. Instead, most systems transformed patient histories into increasingly compressed representations, including aggregated features, latent embeddings, risk scores, recommendations, or generated summaries.

\subsection{Current EHR-AI systems}
The results suggest that much of the current EHR-AI literature is organized around prediction instead of representation. Predictive performance has become a dominant benchmark for evaluating clinical AI systems, leading many architectures to optimize for forecasting outcomes and risk, or generating recommendations. However, prediction and interpretation are not equivalent. Recent work in interpretable machine learning has argued that highly accurate predictions do not necessarily provide explanations that support human reasoning or decision-making, particularly in high-stakes domains such as medicine \citep{Rudin2019}. Likewise, studies of clinician perspectives on explainable AI suggest that physicians seek contextualized representations that preserve clinical reasoning pathways rather than isolated model outputs or feature attributions \citep{Tonekaboni2019}. A risk score may indicate that a patient is likely to deteriorate, but it does not necessarily help clinicians understand the temporal sequence of events that produced that risk, the unresolved clinical questions that remain, or the competing explanations that should be considered. As a result, systems can achieve strong predictive performance while providing limited support for reconstructing the patient's clinical trajectory.

This distinction is reflected in the longitudinal fidelity findings. Only a minority of systems were coded as maintaining high-fidelity representations of longitudinal patient state, while most exhibited low or very low longitudinal fidelity. Figure 3 further demonstrates that high longitudinal fidelity was concentrated almost entirely among systems that explicitly modeled patients across time. In contrast, encounter-level and population-level systems overwhelmingly produced compressed representations of clinical information. These findings suggest that preservation of longitudinal context is not an automatic consequence of using EHR data. Rather, it is a specific architectural and representational choice.

The concept of representational compression provides a useful framework for interpreting these findings. EHRs contain rich, temporally structured records that capture symptoms, diagnoses, treatments, responses to therapy, laboratory trends, clinician observations, and evolving hypotheses over time. Many AI systems consume this information but ultimately expose only a small subset of it through scores, classifications, alerts, recommendations, or summaries. Such compression can improve efficiency and scalability, but it may also obscure clinically meaningful context. Information that is relevant for understanding how a patient's condition evolved can become difficult or impossible to recover once the underlying trajectory has been compressed into a single output.

One notable finding from the review is the relative rarity of explicit reasoning about absence. Clinical interpretation often depends not just on what is present in the record, but on what is absent too. For example, an absence of fever, lack of response to antibiotics, failure of laboratory values to normalize, or missing expected follow-up can substantially alter the interpretation of a case. Yet many systems primarily represent observed events and documented findings, leaving limited mechanisms for incorporating absent or expected-but-missing information into their representations. This suggests a potentially important gap between current EHR-AI architectures and the informational structure of real-world clinical practice.

A related challenge concerns uncertainty. Clinical care unfolds under conditions of incomplete information and evolving evidence. In contrast, many AI systems ultimately produce discrete classifications or fluent summaries that can obscure the uncertainty inherent in the underlying clinical situation. While uncertainty quantification has received increasing attention in machine learning research, the literature reviewed here suggests that uncertainty is often represented less explicitly than predictive outputs themselves. Future systems may benefit from making uncertainty visible components of the patient representation rather than treating them as downstream concerns.

The physician reflections included in this review provide an important complementary perspective. Although the literature frequently emphasizes model performance, clinicians often experience EHRs as tools for reconstructing patient histories and understanding how interpretations have changed over time. The workflow challenges described by physicians align closely with the representational limitations identified in the corpus. Difficulties tracing evolving clinical narratives and recommendations suggest that the representational choices made by AI systems have direct implications for usability and clinical decision-making. Similar observations have emerged from implementation science, where successful deployment of clinical machine learning systems has been shown to depend as much on workflow integration and human-AI interaction as on model performance itself \citep{Sendak2020}.

These findings have several implications for the design of future EHR-integrated AI systems. First, patient trajectories should be treated as primary representational objects rather than incidental inputs to predictive models. Second, AI-generated outputs should preserve explicit links to the underlying evidence from which they were derived, allowing clinicians to inspect and verify supporting information. Third, systems should provide mechanisms for visualizing how assessments change over time rather than presenting only static outputs. Fourth, future architectures should distinguish between observed facts, inferred relationships, missing information, and uncertain conclusions, thereby making the structure of clinical knowledge more transparent.

The findings also suggest opportunities for improving evaluation methodologies. Current benchmarks frequently emphasize predictive accuracy, discrimination, or task completion. This reflects a broader tendency within clinical AI to evaluate algorithmic outputs in isolation, despite growing recognition that meaningful clinical decision support requires systems that interact effectively with human reasoning processes \citep{Shortliffe2018}. While these metrics remain important, they do not capture whether a system preserves clinically meaningful longitudinal context. Future evaluations should consider longitudinal reasoning tasks, trajectory reconstruction, evidence traceability, uncertainty representation, and clinician-centered measures of interpretability and usability. Such evaluations would better reflect the role AI systems play within longitudinal clinical workflows.

Taken together, the results suggest that the next challenge for EHR-integrated AI may not be obtaining more clinical data or achieving marginal improvements in predictive performance. Instead, it may be developing systems that better preserve and expose the temporal structure of patient histories. The next generation of EHR-integrated AI should therefore be evaluated by how effectively it helps clinicians reconstruct and interpret a patient's evolving clinical trajectory. This perspective is consistent with the broader goal of clinical informatics as a discipline: developing computational systems that amplify human expertise by making complex clinical information more interpretable and actionable than simply more computationally tractable \citep{Friedman1997, Shortliffe2018}.

\subsection{Information Compression as a Representational Framework for EHR-Based AI Systems}
Electronic health records have long been recognized as a rich substrate for computational modeling and secondary analysis, enabling large-scale biomedical discovery while requiring substantial transformation of complex longitudinal clinical data into computational representations \citep{Jensen2012}. More recently, advances in deep learning have further accelerated the development of increasingly sophisticated representations of patient information across prediction and clinical decision support tasks \citep{Esteva2019}. Building upon this broader evolution of EHR-based AI, we propose information compression as a unifying representational framework for understanding how contemporary systems transform longitudinal patient histories as they move through computational pipelines. At one end of the continuum are raw retrieval systems, which preserve much of the original patient record and allow clinicians or downstream systems to access underlying observations directly. Examples include large clinical databases and retrieval-oriented infrastructures that primarily organize and expose information and not transform it. These systems maintain the highest degree of longitudinal fidelity because patient trajectories remain largely intact.

Moving further along the continuum, feature aggregation systems transform patient histories into structured variables or extracted features. While these approaches reduce complexity and facilitate statistical modeling, they necessarily discard portions of the original narrative and contextual structure present within longitudinal records. Many traditional clinical prediction systems and natural language processing pipelines occupy this region of the continuum.

Latent embedding systems compress patient histories even further by representing individuals as learned vector representations. Models such as Deep Patient, Doctor AI, BEHRT, CLMBR, and related architectures maintain important longitudinal patterns but do so within highly compressed latent spaces. These representations can preserve clinically meaningful relationships while simultaneously reducing interpretability and direct access to the underlying patient trajectory.

Further compression occurs in recommendation-generating systems, which transform complex patient histories into alerts, risk scores, classifications, or decision support outputs. Although these systems may utilize rich longitudinal information internally, the information ultimately exposed to clinicians is substantially condensed. The output often consists of a small number of recommendations or probability estimates rather than an explicit representation of the patient's evolving clinical state.

At the most compressed end of the continuum are summary-generation systems, such as ambient or automated note generators, and certain large language model applications. These systems convert large volumes of patient information into concise textual outputs intended for human consumption. While such summaries may improve efficiency and usability, they frequently obscure the temporal and relational structure of the underlying patient history.

This framework suggests that the increasing sophistication of AI models does not necessarily correspond to increasing preservation of longitudinal patient information. In many cases, more advanced systems perform more aggressive representational compression. Consequently, access to longitudinal EHR data should not be conflated with maintenance of longitudinal patient state. A system may consume years of patient history while ultimately exposing only a single prediction, recommendation, or summary.

Viewed through this lens, a key challenge for future EHR-based AI systems is not merely improving predictive performance but balancing information compression with preservation of clinically meaningful temporal context. Models that maintain richer representations of patient trajectories may provide a more faithful account of evolving clinical states while still supporting efficient decision-making. Future research should therefore consider longitudinal fidelity as a complementary evaluation dimension alongside accuracy, discrimination, calibration, and usability.

\section{Limitations}
Several limitations should be considered when interpreting these findings. First, although the corpus was systematically assembled and audited, it was not exhaustive. Consequently, the review may not capture every EHR-integrated AI system described in the literature, particularly recently published systems, proprietary implementations, or models described outside traditional academic venues.

Second, the coding framework necessarily involved interpretive judgment. Variables such as longitudinal fidelity, patient representation level, trajectory modeling, and reasoning-related features required qualitative assessment of system descriptions and architectures. While coding criteria were developed to promote consistency, alternative interpretations may be possible for some systems.

Third, the analysis relied on information reported within published papers. As a result, absence of a capability in the coding framework does not necessarily indicate absence of that capability in the underlying system. Certain features, particularly those related to longitudinal reasoning, uncertainty representation, or handling of missing information, may be underreported or insufficiently described in published manuscripts.

Fourth, the physician reflections were based on three physicians and should be interpreted as illustrative and not representative. These narratives were included to provide clinical context and identify recurring workflow challenges, not to serve as a formal qualitative study. In addition, EHR systems are often heavily customized at the institutional level, meaning that physician experiences may vary substantially even when using the same underlying platform.

Fifth, this review focused on representational and methodological characteristics of EHR-integrated AI systems and did not directly compare implemented performances. The findings therefore should not be interpreted as evidence that one representational approach is necessarily superior in predictive accuracy or clinical effectiveness.

Finally, the field is evolving rapidly. The emergence of large language models, foundation models, multimodal architectures, and increasingly interactive AI systems may alter how clinical information is represented and utilized. As future systems become more agentic, conversational, and longitudinally aware, the coding framework introduced in this review may require refinement to capture new forms of representation, interaction, and clinical decision support.

\section{Future Work}
This review introduces a framework for characterizing how EHR-integrated AI systems represent, compress, and preserve longitudinal patient information. However, the framework should be viewed as an initial step rather than a definitive taxonomy. Future work should expand the corpus, incorporate additional publication sources, and perform larger systematic reviews to determine whether the patterns observed here generalize across the rapidly evolving clinical AI landscape. Validation of the coding framework through multiple independent reviewers, larger paper samples, and formal inter-rater reliability analyses would further strengthen its utility as a tool for characterizing EHR-based AI systems.

Beyond expanding the literature base, future research should develop benchmark tasks that more directly evaluate longitudinal clinical reasoning. Existing evaluations frequently focus on prediction, classification, question answering, or documentation tasks, yet these metrics do not assess whether systems preserve clinically meaningful trajectories or support reconstruction of complex patient histories. New benchmarks should evaluate the ability of systems to identify temporal relationships, track evolving diagnoses, reconstruct clinical narratives across encounters, and distinguish between observed findings, inferred conclusions, unresolved questions, and missing information.

A particularly important direction will be evaluating AI systems within clinician workflows and not just retrospective model performance. Future studies should examine whether AI-assisted systems help clinicians reconstruct complex patient histories more rapidly, identify relevant information more accurately, and make more informed decisions under realistic clinical time constraints. Such evaluations should compare conventional approaches based on static summaries and risk scores against more interactive systems that support exploration of longitudinal patient histories and evolving clinical trajectories.

The findings of this review also suggest that future EHR-integrated AI systems may benefit from moving beyond encounter-centric and prediction-centric paradigms. Future systems could maintain dynamic representations of patient state that evolve over time as new information becomes available. Such representations could make temporal relationships, changes in interpretation, treatment responses, and evolving risk more explicit and accessible to clinicians.

Several specific representational challenges remain largely unaddressed in the current literature. Future work should explore methods for representing absent findings, missing information, failed interventions, and unresolved hypotheses, all of which frequently play important roles in clinical decision-making. Additionally, future systems should look into mechanisms to preserve uncertainty and links between generated output and source material, as well as communicate how clinical interpretations change over time.

The physician reflections included in this study further suggest that future research should account for the substantial variability in how longitudinal reasoning occurs across clinical specialties. The informational needs of primary care, emergency medicine, oncology, cardiology, psychiatry, and critical care differ considerably, and systems optimized for one domain may inadequately support another. Specialty-specific studies may therefore be necessary to understand how patient trajectories should be represented and presented in different clinical contexts.

More broadly, this review argues that the next generation of EHR-integrated AI should be evaluated as both predictive systems and representational systems. The central challenge is no longer simply extracting information from the EHR, but organizing that information in ways that support clinical understanding. Future AI systems should help clinicians reconstruct how a patient's condition evolved, why particular decisions were made, what uncertainties remain, and which pieces of evidence support competing interpretations. Achieving this goal will likely require a shift from viewing the EHR as a repository of documents toward viewing it as a dynamic representation of an evolving patient trajectory.

\section{Conclusion}
Electronic health records have become the primary data substrate for contemporary clinical artificial intelligence. As a result, AI systems are increasingly integrated into prediction models, decision support tools, retrieval systems, and generative applications across clinical practice. The findings of this review, however, suggest that widespread access to EHR data alone does not guarantee support for the longitudinal and interpretive processes that characterize clinical reasoning.

By examining a diverse corpus of EHR-integrated AI systems, this paper argues for a distinction between three related but conceptually separate ideas: data access, data representation, and reasoning support. While many contemporary systems successfully access and process longitudinal patient information, comparatively few preserve the evolving temporal and contextual relationships that clinicians rely upon when interpreting patient histories across encounters. Longitudinal fidelity therefore emerges not simply as a technical characteristic of individual models, but as a broader representational property that shapes how clinical information is made available for reasoning.

The physician reflections reinforce the practical importance of this distinction. Despite increasingly comprehensive digital records, clinicians continue to describe substantial effort devoted to identifying relevant information and understanding how prior clinical conclusions were reached. These challenges parallel the representational patterns identified throughout the literature, suggesting that the organization of clinical information may be as important as its availability.

Taken together, these observations point toward a broader shift in how EHR-integrated AI systems should be conceptualized and evaluated. Existing benchmarks have understandably emphasized predictive accuracy, discrimination, and task completion, yet these measures capture only one dimension of system performance. Future evaluation frameworks may also benefit from assessing the extent to which systems preserve longitudinal patient trajectories, maintain evidence traceability, represent uncertainty, support cross-encounter synthesis, and facilitate clinician interpretation over time. Such measures would recognize that AI systems participate in both the prediction and communication of clinical knowledge.

More broadly, this work argues that the next stage of EHR-integrated AI development is fundamentally a representational challenge. As healthcare increasingly adopts AI-enabled workflows, the central question is no longer whether systems can extract information from electronic health records, but whether they can organize and present that information in ways that align with the temporal, cumulative, and interpretive nature of clinical reasoning. Systems that preserve patient trajectories as explicit representational objects may better support clinicians altogether.

Ultimately, realizing the full potential of EHR-integrated AI will require moving beyond viewing the electronic health record as a repository of documents or a source of predictive features. Instead, it should be understood as a dynamic representation of an unfolding clinical story, one that future AI systems should help clinicians interpret rather than merely summarize or predict.

\bibliographystyle{apalike}
\bibliography{sample}

@misc{HITECH2009,
  author       = {HITECH},
  title        = {Health Information Technology for Economic and Clinical Health Act},
  year         = {2009},
  howpublished = {Title XIII of the American Recovery and Reinvestment Act of 2009, Pub. L. No. 111-5, 123 Stat. 226},
  note         = {United States Congress}
}

@article{Blumenthal2010,
  author  = {Blumenthal, David},
  title   = {Launching HITECH},
  journal = {New England Journal of Medicine},
  year    = {2010},
  volume  = {362},
  number  = {5},
  pages   = {382--385},
  doi     = {10.1056/NEJMp0912825}
}

@article{Bates2003,
  author  = {Bates, David W. and Kuperman, Gilad J. and Wang, Shirley and Gandhi, Tejal and Kittler, Alice and Volk, Lisa and Spurr, Christopher and Khorasani, Ramin and Tanasijevic, Mark and Middleton, Blackford},
  title   = {Ten Commandments for Effective Clinical Decision Support: Making the Practice of Evidence-Based Medicine a Reality},
  journal = {Journal of the American Medical Informatics Association},
  year    = {2003},
  volume  = {10},
  number  = {6},
  pages   = {523--530}
}

@article{Kawamoto2005,
  author  = {Kawamoto, Kensaku and Houlihan, C. Andrew and Balas, E. Andrew and Lobach, David F.},
  title   = {Improving Clinical Practice Using Clinical Decision Support Systems: A Systematic Review of Trials to Identify Features Critical to Success},
  journal = {BMJ},
  year    = {2005},
  volume  = {330},
  number  = {7494},
  pages   = {765},
  doi     = {10.1136/bmj.38398.500764.8F}
}

@article{Bright2012,
  author  = {Bright, Tiffani J. and Wong, Angela and Dhurjati, Ritu and Bristow, Elizabeth and Bastian, Lori and Coeytaux, Remy R. and Samsa, Gregory and Hasselblad, Vic and Williams, John W. and Musty, Matthew D. and Wing, Lisa and Kendrick, Amanda S. and Sanders, Gary D. and Lobach, David},
  title   = {Effect of Clinical Decision-Support Systems: A Systematic Review},
  journal = {Annals of Internal Medicine},
  year    = {2012},
  volume  = {157},
  number  = {1},
  pages   = {29--43},
  doi     = {10.7326/0003-4819-157-1-201207030-00450}
}

@inproceedings{Choi2016,
  author    = {Choi, Edward and Bahadori, Mohammad Taha and Schuetz, Andy and Stewart, Walter F. and Sun, Jimeng},
  title     = {Doctor AI: Predicting Clinical Events via Recurrent Neural Networks},
  booktitle = {Proceedings of Machine Learning for Healthcare},
  year      = {2016},
  pages     = {301--318}
}

@article{Rajkomar2018,
  author  = {Rajkomar, Alvin and Oren, Eyal and Chen, Kai and Dai, Andrew M. and Hajaj, Nir and Hardt, Michaela and Liu, Peter J. and Liu, Xiaobing and Marcus, Jake and Sun, Mike and Sundberg, Patrik and Yee, Hector and Zhang, Kun and Duggan, Gavin E. and Irvine, James and Le, Quoc V. and Litsch, Kurt and Dean, Jeff and Socher, Richard},
  title   = {Scalable and Accurate Deep Learning with Electronic Health Records},
  journal = {npj Digital Medicine},
  year    = {2018},
  volume  = {1},
  pages   = {18},
  doi     = {10.1038/s41746-018-0029-1}
}

@article{Shah2019,
  author  = {Shah, Nigam H. and Milstein, Arnold and Bagley, Steven C.},
  title   = {Making Machine Learning Models Clinically Useful},
  journal = {JAMA},
  year    = {2019},
  volume  = {322},
  number  = {14},
  pages   = {1351--1352},
  doi     = {10.1001/jama.2019.10306}
}

@article{Rudin2019,
  author  = {Rudin, Cynthia},
  title   = {Stop Explaining Black Box Machine Learning Models for High Stakes Decisions and Use Interpretable Models Instead},
  journal = {Nature Machine Intelligence},
  year    = {2019},
  volume  = {1},
  number  = {5},
  pages   = {206--215},
  doi     = {10.1038/s42256-019-0048-x}
}

@article{Singhal2023,
  author  = {Singhal, Karan and Azizi, Shekoofeh and Tu, Tao and Mahdavi, Sara S. and Wei, Jason and Chung, Hyung Won and Scales, Nathan and others},
  title   = {Large Language Models Encode Clinical Knowledge},
  journal = {Nature},
  year    = {2023},
  volume  = {620},
  number  = {7972},
  pages   = {172--180},
  doi     = {10.1038/s41586-023-06291-2}
}

@article{DoshiVelez2017,
  author  = {Doshi-Velez, Finale and Kim, Been},
  title   = {Towards a Rigorous Science of Interpretable Machine Learning},
  journal = {arXiv preprint arXiv:1702.08608},
  year    = {2017},
  eprint  = {1702.08608},
  archivePrefix = {arXiv}
}

@article{Chenais2023,
  author  = {Chenais, Guillaume and Busche, Katharina and Bolognini, Damien and others},
  title   = {Artificial Intelligence in Emergency Medicine: Viewpoint of Current Applications and Future Perspectives},
  journal = {Journal of Medical Internet Research},
  year    = {2023},
  volume  = {25},
  pages   = {e40031},
  doi     = {10.2196/40031}
}

@article{Friedman1997,
  author = {Friedman, Charles P.},
  title = {What Informatics Is and Isn't},
  journal = {Journal of the American Medical Informatics Association},
  year = {1997},
  volume = {4},
  number = {3},
  pages = {157--160},
  doi = {10.1136/jamia.1997.0040157}
}

@book{Topol2019,
  author = {Topol, Eric},
  title = {Deep Medicine: How Artificial Intelligence Can Make Healthcare Human Again},
  publisher = {Basic Books},
  year = {2019}
}

@article{Jensen2012,
  author = {Jensen, Peter B. and Jensen, Lars Juhl and Brunak, Søren},
  title = {Mining Electronic Health Records: Towards Better Research Applications and Clinical Care},
  journal = {Nature Reviews Genetics},
  year = {2012},
  volume = {13},
  number = {6},
  pages = {395--405},
  doi = {10.1038/nrg3208}
}

@article{Esteva2019,
  author = {Esteva, Andre and Robicquet, Alexandre and Ramsundar, Bharath and others},
  title = {A Guide to Deep Learning in Healthcare},
  journal = {Nature Medicine},
  year = {2019},
  volume = {25},
  number = {1},
  pages = {24--29},
  doi = {10.1038/s41591-018-0316-z}
}

@article{Sutton2020,
  author = {Sutton, Richard T. and Pincock, David and Baumgart, Daniel C. and others},
  title = {An Overview of Clinical Decision Support Systems: Benefits, Risks, and Strategies for Success},
  journal = {NPJ Digital Medicine},
  year = {2020},
  volume = {3},
  pages = {17},
  doi = {10.1038/s41746-020-0221-y}
}

@article{Tonekaboni2019,
  author = {Tonekaboni, Sana and Joshi, Shalmali and McCradden, Melissa D. and Goldenberg, Anna},
  title = {What Clinicians Want: Contextualizing Explainable Machine Learning for Clinical End Use},
  journal = {Machine Learning for Healthcare Conference},
  year = {2019},
  pages = {359--380}
}

@article{Sendak2020,
  author = {Sendak, Mark P. and D'Arcy, John and Kashyap, Sandeep and others},
  title = {A Path for Translation of Machine Learning Products into Healthcare Delivery},
  journal = {EMJ Innovations},
  year = {2020},
  volume = {4},
  number = {1},
  pages = {38--45}
}

@article{Shortliffe2018,
  author = {Shortliffe, Edward H. and Sepúlveda, Maria J.},
  title = {Clinical Decision Support in the Era of Artificial Intelligence},
  journal = {JAMA},
  year = {2018},
  volume = {320},
  number = {21},
  pages = {2199--2200},
  doi = {10.1001/jama.2018.17163}
}

\end{document}